\documentclass[12pt]{iopart}
\usepackage{iopams}
\usepackage{csquotes}
\usepackage{graphicx}
\usepackage{dcolumn}
\usepackage{bm}
\usepackage{hyperref}
\usepackage[square,sort,comma,numbers]{natbib}

\newtheorem{Remark}{Remark}
\begin{document}

\title[Wigner spectrum and coherent feedback control of continuous-mode single-photon states]{Wigner spectrum and coherent feedback control of continuous-mode single-photon Fock states}

\author{Zhiyuan Dong$^1$, Lei Cui$^1$, Guofeng Zhang$^1$ and Hongchen Fu$^2$}

\address{$^1$ Department of Applied Mathematics, The Hong Kong Polytechnic University, Hong Kong, China}

\address{$^2$ School of Physical Sciences and Technology, Shenzhen University, Shenzhen 518060, China}
\ead{Guofeng.Zhang@polyu.edu.hk}
\vspace{10pt}
\begin{indented}
\item[]July 2016
\end{indented}

\begin{abstract}
Single photons are very useful resources in quantum information science. In real applications it is often required that the photons have a well-defined spectral (or equivalently temporal) modal structure. For example, a rising exponential pulse is able to  fully excite a two-level atom while a Gaussian pulse cannot. This motivates the study of continuous-mode single-photon Fock states. Such states are characterized by a spectral (or temporal) pulse shape. In this paper we investigate the statistical property of  continuous-mode single-photon Fock states. Instead of the commonly used normal ordering (Wick order), the tool we proposed is the Wigner spectrum. The Wigner spectrum has two advantages: 1) it allows to study continuous-mode single-photon Fock states in the time domain and frequency  domain simultaneously; 2) because it can deal with the Dirac delta function directly, it has the potential to provide more information than the normal ordering where the Dirac delta function is always discarded.  We also show how various control methods in particular coherent feedback control can be used to manipulate the pulse shapes of continuous-mode single-photon Fock states.
\end{abstract}

%
%
\submitto{\jpa}
%
%
%
(Some figures may appear in color only in the online journal)

\section{Introduction}

Single photons are fundamental resources for quantum communication \cite{beveratos2002single,gisin2007quantum}, quantum computing \cite[Chapter 6.3]{loudon2000quantum}, \cite{knill2001scheme}, quantum metrology \cite{giovannetti2004quantum,giovannetti2011advances,leroux2012unitary}, and quantum networks \cite{moehring2007entanglement,kimble2008quantum,aghamalyan2011quantum}. In contrast to single-mode photon states, continuous-mode photon states are closer to a real experimental environment in quantum information processing \cite{blow1990continuum,braunstein2005quantum,furusawa2011quantum,brecht2015photon}. Continuous-mode photon states are characterized by a well-defined temporal (or equivalently spectral) modal structure, often called pulse shape \cite{blow1990continuum, loudon2000quantum,garrison2008quantum,milburn2008coherent,ogawa2015real}.   In \cite{WMS+11}, the authors discussed efficient excitation of a two-level atom by a continuous-mode single-photon Fock state. The effect of various temporal pulse shapes (for example Gaussian, hyperbolic secant, rectangular, rising exponential and decaying exponential) on the excitation probability is studied.  Recently, an experiment has been conducted which demonstrated real-time measurement of  a rising exponential single-photon wavepacket \cite{ogawa2015real}. Quantum filters for an arbitrary quantum system driven by a continuous-mode single-photon Fock state has been investigated in \cite{gough2011quantum,gough2012quantum,gough2013quantum}. The study in \cite{gough2012quantum,gough2013quantum} are extended in \cite{baragiola2012n} to derive the master equations of an arbitrary quantum system driven by continuous-mode multi-photon Fock wave packets. Lately, based on \cite{gough2013quantum}, the quantum filters of an arbitrary quantum system driven by a continuous-mode multi-photon state are derived in \cite{SZX16}. By applying the stochastic master equations to a cavity driven by a continuous-mode single-photon field, the conditional dynamics for the cross phase modulation in a doubly resonant cavity are analyzed in \cite{carvalho2012cavity}.

The statistical properties of continuous-mode single-photon Fock states have been studied, see e.g.,  \cite{gheri1998photon,loudon2000quantum,milburn2008coherent,zhang2013response,yamamoto2014zero,qin2014complete}.  In most of these studies normal ordering is often used. Let $\hat{b}(t)$ be a boson annihilation operator of a travelling field, the normal ordering of the product $\hat{b}(t)\hat{b}^{\dagger }(r)$ is
$:\hat{b}(t)\hat{b}^{\dagger }(r):=\hat{b}^{\dagger }(r)\hat{b}(t).$
That is, the Dirac delta function  $\delta (t-r)$ has been thrown away.  As a result, partial information has been lost in the procedure of normal ordering. In this paper we propose an alternative method, the Wigner spectrum, to study the statistical property of continuous-mode single-photon Fock states. We show that the Wigner spectrum can handle the Dirac delta function naturally, thus no information is abandoned.  Moreover, the Wigner spectrum allows us to visualize continuous-mode single-photon Fock states in both the time domain and frequency domain simultaneously.

In the input-output formalism, the problem of pulse-shaping of continuous-mode single-photon Fock states has been investigated in \cite{milburn2008coherent}. The relation between input and output pulse shapes is derived in the frequency domain when the underlying system is an empty cavity.  Based on the cross-phase shift of a coherent state induced by a single-photon state, a weak nonlinearity phase gate is discussed in \cite{munro2010intracavity}. The input-output relation of pulse shapes is expressed by transfer function in \cite{zhang2013response}. The pulse-shaping problem in the case of quantum linear systems has also been discussed. A memory subsystem within a linear network is proposed in \cite{yamamoto2014zero}.  The response of quantum nonlinear systems to single-photon input states is presented in \cite{pan2014input}. Particularly, the output states and pulse shapes for quantum two-level systems are derived explicitly in the time and frequency domains. In this paper, we demonstrate how various control methods (direct coupling and coherent feedback control) can be used for pulse-shaping of continuous-mode single-photon Fock states. The effect of control techniques on pulse-shaping is visualized by the Wigner spectrum of the output single-photon states.

\section{Wigner spectrum for optical cavity}\label{Wig_cavity}

\subsection{Single-photon states}

In this paper we study quasi-monochramatic light fields. Such a light field has a frequency profile around its carrier (central) frequency. The bandwidth of the profile is much smaller than the carrier frequency.    Let $\hat{b}[\omega ]$ be the boson annihilation operator for mode $\omega $ of
the light field. $\hat{b}[\omega ]$ and its adjoint operator $\hat{b}%
^{\dagger }[\omega ]$ satisfy the singular commutation relation
\begin{equation}
\left[ \hat{b}[\omega _{1}],\hat{b}^{\dagger }[\omega _{2}]\right] =\delta
\left( \omega _{1}-\omega _{2}\right) .  \label{june29_CCR}
\end{equation}%
Define an operator in the interaction picture
\begin{equation}
\mathbf{\hat{B}}(\xi )\triangleq \int_{-\infty }^{\infty }d\omega \ \xi^{\ast}
[\omega ]\hat{b}[\omega ]  \label{B}
\end{equation}%
with a normalized spectral pulse shape $\xi
\lbrack \omega ]$, i.e.,  $\left\Vert \xi \right\Vert \triangleq \int_{-\infty }^{\infty
}d\omega \ \left\vert \xi \lbrack \omega ]\right\vert ^{2}=1$. A \textit{%
continuous-mode single-photon Fock state} is defined to be
\begin{equation}
\left\vert 1_{\xi }\right\rangle \equiv \mathbf{\hat{B}}^{\dagger }(\xi
)\left\vert 0\right\rangle \triangleq \int_{-\infty }^{\infty }d\omega \ \xi
\lbrack \omega ]\hat{b}^{\dagger }[\omega ]\left\vert 0\right\rangle .
\label{eq:1_photon}
\end{equation}%
$\hat{b}^{\dagger }[\omega ]$ is the creation operator of the light field,
and thus $\hat{b}^{\dagger }[\omega ]\left\vert 0\right\rangle \equiv
\left\vert 1_{\omega }\right\rangle $ can be understood as photon generation
at frequency $\omega $, while the probability is given by $\left\vert \xi
\lbrack \omega ]\right\vert ^{2}$. So the continuous-mode single-photon Fock
state $\left\vert 1_{\xi }\right\rangle $ can be interpreted as a photon
coherently superposed over a continuum of frequency modes, with probability
amplitudes given by the spectral density function $\xi \lbrack \omega ]$.
The Fourier transform of \eref{eq:1_photon} gives the time domain
expression of the single-photon Fock state, which is
\begin{equation}
\left\vert 1_{\xi }\right\rangle =\int_{-\infty }^{\infty }dt\ \xi (t)\hat{b}%
^{\dagger }(t)\left\vert 0\right\rangle .  \label{eq:1_photon_t}
\end{equation}%
Clearly, the time-domain counterpart of the commutation relation (\ref%
{june29_CCR}) is%
\begin{equation}
\left[ \hat{b}(t),\hat{b}^{\dagger }(r)\right] =\delta \left( t-r\right) .
\label{june30_CCR}
\end{equation}%
It is easy to show that for the continuous-mode single-photon Fock state $%
\left\vert 1_{\xi }\right\rangle $, the average field amplitude is zero, that is,
\begin{equation}
\left\langle 1_{\xi }|\mathbf{\hat{B}}^{\dagger }(\xi )|1_{\xi
}\right\rangle =\left\langle 1_{\xi }|\mathbf{\hat{B}}(\xi )|1_{\xi
}\right\rangle =0.  \label{june29_2a}
\end{equation}%
Moreover,%
\begin{equation}
\mathbf{\hat{B}}(\xi )\left\vert 1_{\xi }\right\rangle =\left\Vert \xi
\right\Vert ^{2} |0\rangle=|0\rangle.  \label{june29_2b}
\end{equation}

Next we discuss \textit{continuous-mode single-photon coherent states}, which
can be defined to be, \cite[Eq. (3.1)]{blow1990continuum}
\numparts\begin{eqnarray}
\left\vert \alpha _{\xi }\right\rangle  &=&\exp \left(\alpha \mathbf{\hat{B}}(\xi
)^{\dagger }-\alpha ^{\ast }\mathbf{\hat{B}}(\xi)\right)\left\vert 0\right\rangle
\label{june29_coherent_1a} \\
&=&\exp \left(\int_{-\infty }^{\infty }d\omega \ \alpha \xi \lbrack \omega ]\hat{b%
}^{\dagger }[\omega ]-\int_{-\infty }^{\infty }d\omega \ (\alpha \xi \lbrack
\omega ])^{\ast }\hat{b}[\omega ]\right)\left\vert 0\right\rangle,
\label{june29_coherent_1b}
\end{eqnarray}\endnumparts
where $\alpha =e^{i\theta }$ is a complex number. By the
Baker-Hausdorff formula, $\left\vert \alpha _{\xi }\right\rangle $ can be
re-written as
\begin{equation}
\left\vert \alpha _{\xi }\right\rangle =\exp \left(-\frac{\left\vert \alpha
\right\vert ^{2}}{2}\right)\exp (\alpha \mathbf{\hat{B}}^{\dagger }(\xi )))\exp
(-\alpha ^{\ast }\mathbf{\hat{B}}(\xi ))\left\vert 0\right\rangle .
\label{june29_coherent_2}
\end{equation}%
Consequently, we may express the continuous-mode single-photon coherent
state in terms of continuous-mode number states, that is,

\begin{equation}
\left\vert \alpha _{\xi }\right\rangle =\exp \left(-\frac{\left\vert \alpha
\right\vert ^{2}}{2}\right)\sum_{n=0}^{\infty }\frac{\alpha ^{n}}{\sqrt{n!}}%
\left\vert n_{\xi }\right\rangle ,  \label{june29_coherent_3}
\end{equation}%
where
\begin{equation}
\left\vert n_{\xi }\right\rangle \triangleq \frac{1}{\sqrt{n!}}(\mathbf{\hat{%
B}}^{\dagger }(\xi )^{\dagger })^{n}\left\vert 0\right\rangle   \label{n_xi}
\end{equation}%
is a continuous-mode number state. \eref{june29_coherent_3} is similar
to Eq. (4.3.1) in \cite{gardiner2004quantum}, with the exception of replacing the bosonic
single-mode annihilation operator $\hat{a}$ with the continuous-mode
operator  $\mathbf{\hat{B}}(\xi )$ and accordingly $\left\vert
n\right\rangle $ with $\left\vert n_{\xi }\right\rangle $.

\bigskip It is easy to show that the continuous-mode single-photon coherent
state $\left\vert \alpha _{\xi }\right\rangle $ is the eigenstate of $%
\mathbf{\hat{B}}(\xi )$, that is%
\begin{equation}
\mathbf{\hat{B}}(\xi )\left\vert \alpha _{\xi }\right\rangle =\alpha
\left\vert \alpha _{\xi }\right\rangle .  \label{june29_3b}
\end{equation}%
Moreover,%
\begin{equation}
\left\langle \alpha _{\xi }|\mathbf{\hat{B}}(\xi )|\alpha _{\xi
}\right\rangle =\left\langle \alpha _{\xi }|\mathbf{\hat{B}}^{\dagger }(\xi
)|\alpha _{\xi }\right\rangle =\alpha \left\langle \alpha _{\xi }|\alpha
_{\xi }\right\rangle =\alpha .  \label{june29_3a}
\end{equation}%
And\ the mean photon number is%
\begin{equation}
\left\langle \alpha _{\xi }|\mathbf{\hat{B}}^{\dagger }(\xi )\mathbf{\hat{B}}%
(\xi )|\alpha _{\xi }\right\rangle =\left\vert \alpha \right\vert ^{2}=1.
\label{june29_3c}
\end{equation}%
This is the reason why $\left\vert \alpha _{\xi }\right\rangle $ is called a
\textit{single-photon} coherent state. \

Notice that for any function $\mu \lbrack \omega ]$,

\begin{equation}
\mathbb{E}_{\alpha _{\xi }}\left[ e^{i\int_{-\infty }^{\infty }d\omega \ \mu
\lbrack \omega ]\hat{b}^{\dagger }[\omega ]+\mu ^{\ast }[\omega ]\hat{b}%
[\omega ]}\right] =\mathrm{exp}\left[ -\frac{1}{2}\left\Vert \mu \right\Vert
^{2}+i\left( \left\langle \eta ^{\ast }|\mu \right\rangle +\left\langle \eta
|\mu ^{\ast }\right\rangle \right) \right] ,  \label{june29_4}
\end{equation}%
where $\eta \triangleq \alpha \xi $ and the subscript \textquotedblleft $%
\alpha _{\xi }$\textquotedblright\ indicates that the expectation is taken
with respect to $\left\vert \alpha _{\xi }\right\rangle $. Thus, by the
characteristic function theory, \ $|\alpha _{\xi }\rangle $ is a Gaussian
state. More discussions on continuous-mode coherent states can be found in,
e.g., \cite[Eq. (3.1)]{blow1990continuum}, \cite{gough2005quantum}, and \cite[Section II.E]{zhang2013response}. It should be emphasized
that the Mandel's $Q$ parameters for single-photon Fock state and coherent state
are different. The Mandel's $Q$ parameter for Fock state is less than $0$, which
indicates the sub-Poissonian statistics. While coherent states have a Poissonian
photon-number statistics for which $Q=0$, \cite{blow1990continuum}.

\begin{Remark}
In fact, for the continuous-mode single-photon coherent state, $\alpha \xi
\lbrack \omega ]=e^{i\theta }\xi \lbrack \omega ]$ plays the same role as $%
\alpha (\omega )$ in \cite[Eq. (3.1)]{blow1990continuum}. For the continuous-mode
single-photon Fock state $\left\vert 1_{\xi }\right\rangle$, \eref{eq:1_photon} is
also defined in Section III-B in \cite[Eq. (3.1)]{blow1990continuum}, \cite[Eq. (3)]{gheri1998photon}, \cite[Chapter 6]{loudon2000quantum}, \cite[Eq. (9)]{milburn2008coherent},
\cite[Chapter 5]{garrison2008quantum}, \cite[Eq. (19)]{WMS+11}, \cite[Eq. (17)]{gough2012quantum}, \cite[Eq. (34)]{zhang2013response}.
\end{Remark}

\subsection{Wigner distribution function and Wigner spectrum}

Due to the singular commutation relations \eref{june29_CCR} and
\eref{june30_CCR}, for the continuous-mode single-photon Fock state $%
\left\vert 1_{\xi }\right\rangle $, we have
\begin{equation}
\left\langle 1_{\xi }|\hat{b}(t)\hat{b}^{\dagger }(\tau )|1_{\xi
}\right\rangle =\delta (t-\tau )+\xi (t)\xi^{\ast}(\tau).
\label{june30_1}
\end{equation}%
\eref{june30_1} shows the non-stationarity of the single-photon state $%
\left\vert 1_{\xi }\right\rangle $. The presence of the Dirac delta function
is cumbersome for the statistical analysis of the single-photon state $%
\left\vert 1_{\xi }\right\rangle $. So \textit{normal ordering} is often
used. For example, the normal ordering of $\hat{b}(t)\hat{b}^{\dagger }(r)$
is
\begin{equation}
:\hat{b}(t)\hat{b}^{\dagger }(\tau ):=\hat{b}^{\dagger }(\tau )\hat{b}(t).
\label{june30_2}
\end{equation}%
Notice that in this case,%
\begin{equation}
\left\langle 1_{\xi }|:\hat{b}(t)\hat{b}^{\dagger }(\tau ):|1_{\xi
}\right\rangle =\xi (t)\xi^{\ast}(\tau).  \label{june30_1b}
\end{equation}%
That is, the Dirac delta function is removed. Because of this, time ordering
is commonly used in quantum optics, see e.g., \cite{gardiner2004quantum}. \ In this paper, we
adopt an alternative method for analyzing the statistical properties of
input and output quantum signals. The method we use belongs to the
time-frequency analysis. Let $x(t)$ be a quantum variable, e.g., $\hat{b%
}(t)$, $\hat{b}^{\dagger }(t)$ or $\hat{b}(t)\hat{b}^{\dagger }(t)$, define
the two-time autocorrelation function
\begin{equation}
r_{x}(t,\tau )\triangleq \mathbb{E}_{\xi }[x(t)x^{\dagger }(\tau )],
\label{june30_3}
\end{equation}%
where the subscript \textquotedblleft $\xi $\textquotedblright\ indicates
that the expectation is taken with respect to the single-photon\ state $%
\left\vert 1_{\xi }\right\rangle $.  Clearly, by \eref{june30_1} we
have
\begin{equation}
r_{\hat{b}}(t,\tau )=\mathbb{E}_{\xi }[\hat{b}(t)\hat{b}^{\dagger }(\tau
)]=\delta (t-\tau )+\xi (t)\xi^{\ast}(\tau).  \label{june30_4}
\end{equation}%
Similarly, by normal ordering,
\begin{equation}
r_{\hat{b}^{\dagger }}(\tau ,t)=\mathbb{E}_{\xi }[\hat{b}^{\dagger }(\tau )%
\hat{b}(t)]=\xi (t)\xi^{\ast}(\tau )=\mathbb{E}_{\xi }[:\hat{b}(t)\hat{b}%
^{\dagger }(\tau ):].  \label{june30_4b}
\end{equation}%
Applying the Fourier transform to the two-time autocorrelation function $%
r_{x}(t,\tau )$ with respect to the time variable $\tau$, yields
\begin{equation}
S_{x}(t,\omega )=\frac{1}{\sqrt{2\pi}}\int_{-\infty }^{\infty }r_{x}(t,\tau
)e^{-i\omega \tau }d\tau .  \label{june30_S}
\end{equation}
Define
\begin{equation}
W_{x}(t,\omega )\triangleq \frac{1}{\sqrt{2\pi }}\int_{-\infty }^{\infty
}x(t)x^{\dagger }(\tau )e^{-i\omega \tau }d\tau .  \label{June30_W}
\end{equation}%
Clearly, by \eref{june30_3}, \eref{june30_S}, and \eref{June30_W} we have%
\begin{equation}
S_{x}(t,\omega )=\mathbb{E}_{\xi }\left[ W_{x}(t,\omega )\right] .
\label{June30_SW}
\end{equation}%
In the literature,  $W_{x}(t,\omega )$ is called the \textit{Wigner-Ville
distribution function,} or simply \textit{Wigner function}, and accordingly $%
S_{x}(t,\omega )$  the \textit{Wigner spectrum}, \cite{wigner1932quantum}, \cite{ville1948theorie},
\cite{sandsten2013time}.
Notice that
\begin{equation}
S_{\hat{b}}(t,\omega )=\frac{1}{\sqrt{2\pi }}e^{-i\omega t}+\xi (t)\xi^{\ast}
[\omega].  \label{june30_5a}
\end{equation}%
Comparing \eref{june30_4} and \eref{june30_5a}, we see that the
Dirac delta function does not appear in the Wigner spectrum $S_{x}(t,\omega )
$. Motivated by this, in this paper we use Wigner spectrum to analyze the
statistical properties of quantum signals, instead of resorting to normal
ordering.

\subsection{Optical cavity}

\begin{figure}
\includegraphics[width=0.4\textwidth]{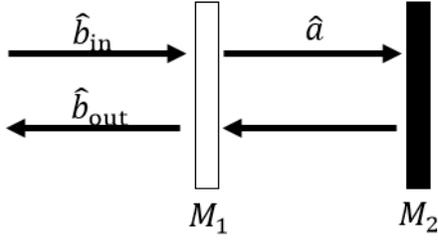}
\caption{\label{cavity} A Fabry-Perot cavity. Arrows indicate the direction of light in the cavity.
Black rectangle ($M_2$) denotes a fully reflecting mirror, while white rectangle ($M_1$) denotes a partially transmitting mirror. $\hat{a}$ is the cavity mode, $\hat{b}_{\rm in}$ is the incident light, and $\hat{b}_{\rm out}$ is the output light.}
\end{figure}

An optical cavity is a system which consists of totally reflecting and/or partially transmitting mirrors \cite{yanagisawa2003transfer}, \cite[Chapter 5.3]{bachor2004guide}, \cite[Chapter 7]{walls2007quantum}, \cite{nurdin2009network}. A widely used type of optical cavities is the so-called Fabry-Perot cavity, as depicted in Fig. \ref{cavity}. In this figure, the electromagnetic filed inside the cavity is mathematically modelled by the bosonic annihilation operator $\hat{a}$. The right-hand mirror ($M_2$) is totally reflecting, while the left-hand mirror ($M_1$) is partially transmitting. The left-hand mirror ($M_1$) allows the incident light (denoted by its annihilation operator $\hat{b}_{\mathrm{in}}$) to enter into the cavity. After bouncing inside the cavity for a while, the electromagnetic field leaves the cavity from the partially transmitting mirror $M_1$, and together with the directly reflected light, forms the outgoing electromagnetic field, as represented by $\hat{b}_{\mathrm{out}}$ in Fig. \ref{cavity}. The coupling between the cavity and the external electromagnetic field is denoted by $\kappa>0$. Moreover, let the de-tuning between the cavity mode and the carrier frequency of the incident light field be $\omega_0$, the dynamics of the Fabry-Parot cavity is, \cite[Chapter 5.3]{gardiner2004quantum}, \cite[Chapter 7]{walls2007quantum}, \cite[Section III]{yanagisawa2003transfer},
\numparts\begin{eqnarray}
\dot{\hat{a}}(t)&=&-(\frac{\kappa}{2}+i\omega_0)\hat{a}(t)-\sqrt{\kappa}\hat{b}_{\mathrm{in}}(t),\\
\hat{b}_{\mathrm{out}}(t)&=&\sqrt{\kappa}\hat{a}(t)+\hat{b}_{\mathrm{in}}(t).
\end{eqnarray}\endnumparts
The impulse response function for system $G$ is given by
\begin{equation}
g_G(t)=\delta(t)-\kappa e^{(-\frac{\kappa}{2}-i\omega_0)t}, ~~ t\geq 0 ,
\end{equation}
while $g_G(t)\equiv 0$ when $t<0$.
Let $|1_\nu\rangle$ be a continuous-mode single-photon Fock state
\begin{equation}
|1_\nu\rangle \equiv {\bf \hat{B}}^\dag(\nu)|0\rangle :=\int_{-\infty}^\infty b_{\mathrm{in}}^\dag (t)\nu(t)dt|0\rangle
\end{equation}
with an exponentially decaying pulse shape
\begin{equation}\label{31}
\nu(t)=\left\{\begin{array}{cc}
                \sqrt{2\gamma}e^{-\gamma t}, & t\geq0, \\
                0, & t<0.
              \end{array}\right.
\end{equation}
The state $|1_\nu\rangle$ can describe a single-photon field emitted from an optical cavity with damping rate $\sqrt{2\gamma}$ \cite{walls2007quantum,loudon2000quantum}. Then the input covariance function is
\begin{eqnarray}
R_{\rm{in}}(t,r)
&\triangleq&
 \mathbb{E}_\nu
\left[
                              \begin{array}{cc}
                                \hat{b}(t)\hat{b}^\dag (r) & \hat{b}(t)\hat{b} (r)  \\
                                \hat{b}^\dag(t)\hat{b}^\dag (r)  & \hat{b}^\dag(t)\hat{b} (r) \\
                              \end{array}
\right]
\nonumber\\
&=&
 \left[
              \begin{array}{cc}
                \delta(t-r) & 0 \\
                0 & 0 \\
              \end{array}
            \right]+\left[
                              \begin{array}{cc}
                                \nu^\ast(r)\nu(t) & 0 \\
                                0 & \nu^\ast(t)\nu(r) \\
                              \end{array}
                            \right].\label{32}
\end{eqnarray}
On the other hand, by the input-output relation \cite{zhang2013response}, the steady-state output single-photon state $|1_\eta\rangle$ has the pulse shape
\begin{equation}\label{33}
\begin{array}{c}
  \eta(t)=\sqrt{2\gamma}e^{-\gamma t}-\displaystyle\frac{\kappa\sqrt{2\gamma}}{\frac{\kappa}{2}+i\omega_0-\gamma}\left(e^{-\gamma t}-e^{(-\frac{\kappa}{2}-i\omega_0)t}\right).
  \end{array}
\end{equation}
The steady-state output covariance function is
\begin{equation}\label{36}
R_{\rm{out}}(t,r)=\delta(t-r)\left[
                        \begin{array}{cc}
                          1 & 0 \\
                          0 & 0 \\
                        \end{array}
                      \right]+\left[
                  \begin{array}{cc}
                  \eta(t)\eta^\ast(r) & 0 \\
                  0                           & \eta^\ast(t)\eta(r)
                  \end{array}
                  \right].
\end{equation}
By \eref{june30_S} and \eref{32}, the Wigner spectrum of the input covariance function can be expressed in terms of both time and frequency
\begin{equation}\label{37}
S_{\rm{in}}(t,\omega)=\frac{1}{\sqrt{2\pi}}\left[
              \begin{array}{cc}
                e^{-i\omega t} & 0 \\
                0 & 0 \\
              \end{array}
            \right]+\frac{1}{\sqrt{2\pi}}\left[
                      \begin{array}{cc}
                        \frac{2\gamma}{\gamma+i\omega}e^{-\gamma t} & 0 \\
                        0 & \frac{2\gamma}{\gamma+i\omega}e^{-\gamma t} \\
                      \end{array}
                    \right].
\end{equation}
Similarly, by \eref{june30_S} and \eref{36}, we can get the Wigner spectrum of the output covariance function
\begin{equation}\label{38}
S_{\rm{out}}(t,\omega)=\frac{1}{\sqrt{2\pi}}\left[
              \begin{array}{cc}
                e^{-i\omega t} & 0 \\
                0 & 0 \\
              \end{array}
            \right]+\frac{1}{\sqrt{2\pi}}\left[
                      \begin{array}{cc}
                        \eta(t)S_{11}[\omega] & 0 \\
                        0 & \eta^\ast(t) S_{22}[\omega] \\
                      \end{array}
                    \right],
\end{equation}
where
\numparts\begin{eqnarray}
S_{11}[\omega]=\sqrt{2\gamma}\times\frac{-\frac{1}{4}\kappa^2+\frac{1}{2}\kappa\gamma-\omega_0^2+\omega\omega_0+i[\gamma\omega_0+\frac{1}{2}\omega\kappa-\omega\gamma]}{(\gamma+i\omega)(\frac{\kappa}{2}-i\omega_0-\gamma)(\frac{\kappa}{2}-i\omega_0+i\omega)},\\
S_{22}[\omega]=\sqrt{2\gamma}\times\frac{-\frac{1}{4}\kappa^2+\frac{1}{2}\kappa\gamma-\omega_0^2-\omega\omega_0+i[-\gamma\omega_0+\frac{1}{2}\omega\kappa-\omega\gamma]}{(\gamma+i\omega)(\frac{\kappa}{2}+i\omega_0-\gamma)(\frac{\kappa}{2}+i\omega_0+i\omega)}.
\end{eqnarray}\endnumparts

If we let decay rate $\kappa\rightarrow\infty$, then the following equation holds
\begin{equation}\label{41}
S_{\rm{out}}(t,\omega)=S_{\rm{in}}(t,\omega).
\end{equation}
That is, the output single-photon state is identical to the input single-photon state.

\begin{figure}
\includegraphics[width=0.8\textwidth]{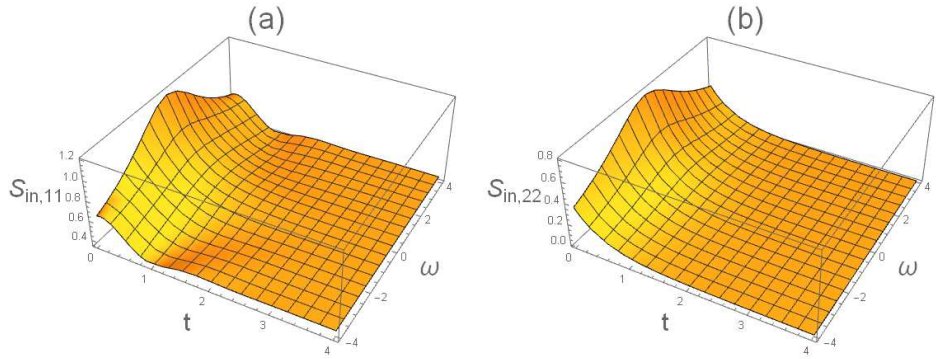}
\caption{\label{fig_15}(Color online) (a) and (b) are the diagonal entries of the input Wigner spectrum.}
\end{figure}

\begin{figure}
\includegraphics[width=0.8\textwidth]{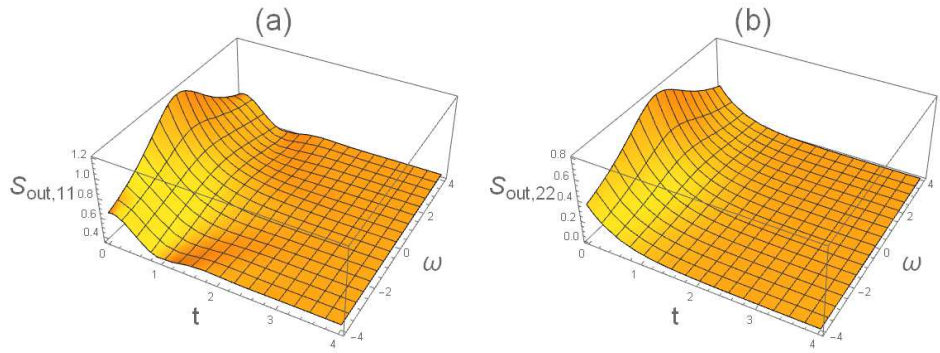}
\caption{\label{fig_16}(Color online) (a) and (b) are the diagonal entries of the output Wigner spectrum with de-tuning $\omega_0=0$ and decay rate $\kappa=0$. The output Wigner spectrum is as same as the input since the output covariance function reduces to the input.}
\end{figure}

\begin{figure}
\includegraphics[width=0.8\textwidth]{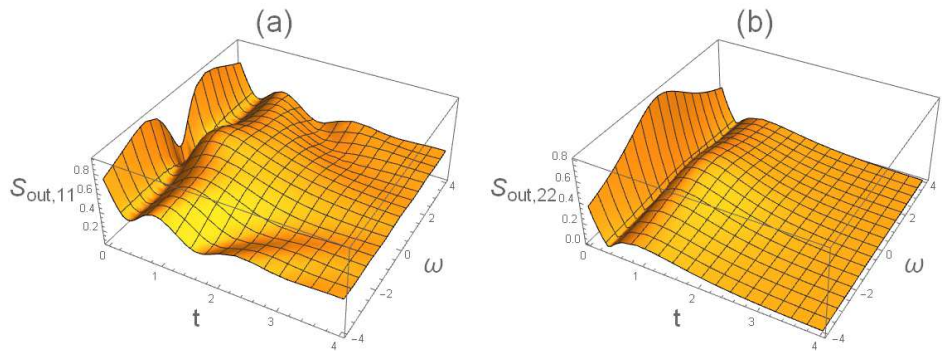}
\caption{\label{fig_19}(Color online) (a) and (b) are the diagonal entries of the output Wigner spectrum with de-tuning $\omega_0=0$ and decay rate $\kappa=3$. Compared with the input, output Wigner spectrum is no longer monotonic in $\omega=0$ since decay rate $\kappa$ becomes larger.}
\end{figure}

\begin{figure}
\includegraphics[width=0.8\textwidth]{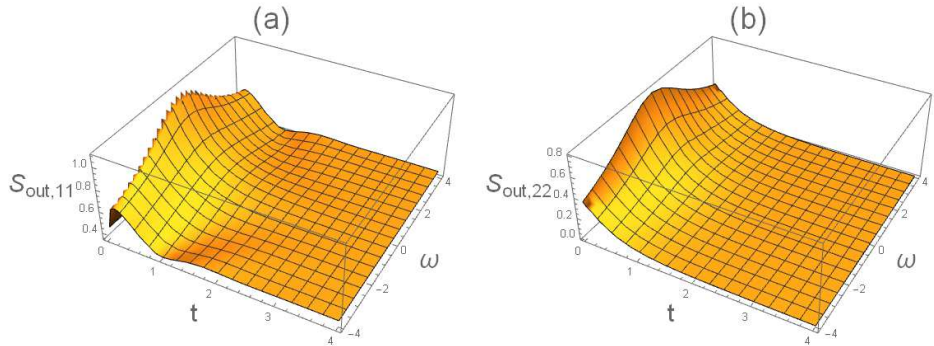}
\caption{\label{fig_20}(Color online) (a) and (b) are the diagonal entries of the output Wigner spectrum with de-tuning $\omega_0=0$ and decay rate $\kappa=100$. The output Wigner spectrum is much similar to the input when decay rate $\kappa$ is large.}
\end{figure}

\begin{figure}
\includegraphics[width=0.8\textwidth]{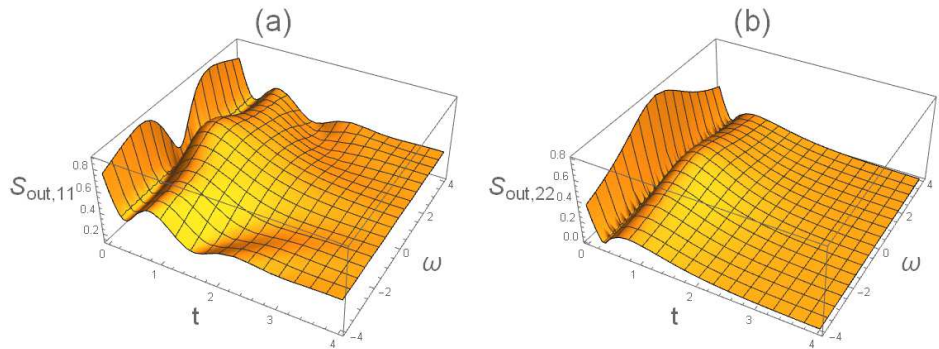}
\caption{\label{fig_21}(Color online) (a) and (b) are the diagonal entries of the output Wigner spectrum with decay rate $\kappa=4$ and de-tuning $\omega_0=0$. In contrast to the decay rate $\kappa$, the output Wigner spectrum is much unlike the input even de-tuning $\omega_0$ is very small.}
\end{figure}

\begin{figure}
\includegraphics[width=0.8\textwidth]{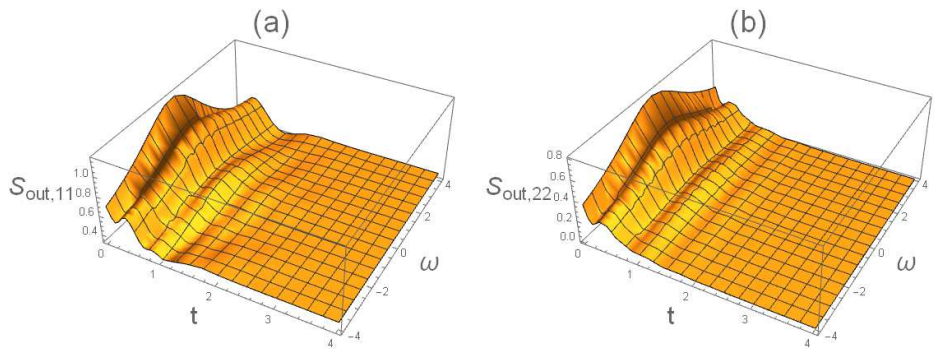}
\caption{\label{fig_23}(Color online) (a) and (b) are the diagonal entries of the output Wigner spectrum with decay rate $\kappa=4$ and de-tuning $\omega_0=10$. When de-tuning $\omega_0$ becomes larger, the output Wigner spectrum will tend to be the input.}
\end{figure}

\begin{figure}
\includegraphics[width=0.8\textwidth]{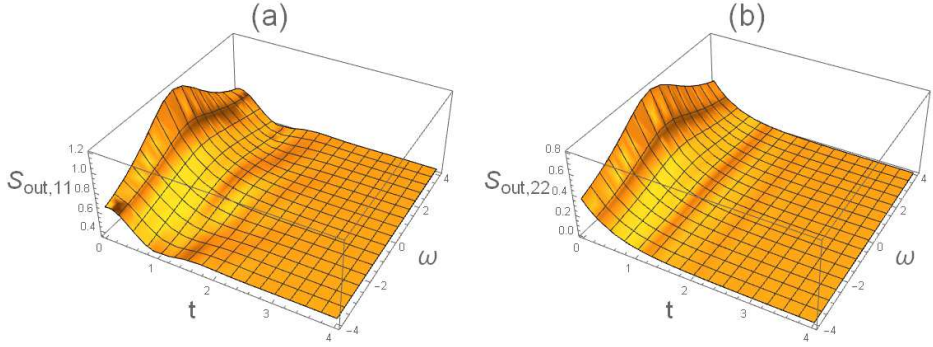}
\caption{\label{fig_25}(Color online) (a) and (b) are the diagonal entries of the output Wigner spectrum with decay rate $\kappa=4$ and de-tuning $\omega_0=50$. Finally, if de-tuning $\omega_0$ is sufficiently large, the output Wigner spectrum would be close to the input.}
\end{figure}

It should be noted that throughout the paper the quantities plotted are  all {\it dimensionless}. In the following we fix damping rate $\gamma=2$. In Fig.~\ref{fig_15}, (a) and (b) are the diagonal entries of the input Wigner spectrum respectively and both of them are exponentially decaying with respect to  time $t$. Fig.~\ref{fig_16}, Fig.~\ref{fig_19} and Fig.~\ref{fig_20} are the output Wigner spectra with different decay rates $\kappa$ and the same de-tuning $\omega_0=0$. Fig.~\ref{fig_21}, Fig.~\ref{fig_23} and Fig.~\ref{fig_25} are the output Wigner spectra with same decay rate $\kappa=4$ and different de-tunings $\omega_0$.

By comparing these figures, we can see that there exist five cases. $\mathbf{Case~1}$: the output Wigner spectrum will be close to the input when the decay rate $\kappa$ is very small (compare Fig. \ref{fig_15} and Fig. \ref{fig_16}); this can be explained by comparing \eref{37} and \eref{38} directly. $\mathbf{Case~2}$: the output Wigner spectrum will also be close to the input when the decay rate $\kappa$ is very large (compare Fig. \ref{fig_15}  and Fig. \ref{fig_20}). Since the impulse response function $g_G(t)\rightarrow\delta(t)$ when $\kappa\rightarrow\infty$, the output state will be close to the input state. $\mathbf{Case~3}$: the output Wigner spectrum would be much similar to the input when the de-tuning $\omega_0$ is very large since the optical cavity has little influence on the photons, see Fig.~\ref{fig_25}. $\mathbf{Case~4}$: it can be seen from Fig.~\ref{fig_19} that the output Wigner spectrum is quite different from the input one when $\kappa$ is not very large or small. Moreover, (a) (for $\hat{b}_{\rm{out}}\hat{b}_{\rm{ out}}^\dag$) and (b) (for $\hat{b}_{\rm{out}}^\dag \hat{b}_{\rm{out}}$) are quite different.   $\mathbf{Case~5}$: The output Wigner spectrum would change a lot with a small de-tuning since there exists a strong interaction between the photon and system (compare Figs. \ref{fig_15} and \ref{fig_21}). Therefore, with Wigner spectrum, we are able to observe the changes of the system's response to the input signals in the time and frequency domains simultaneously. To the best knowledge of the authors, this has not been done before in the single-photon setting.

\section{\label{Wig_DPA}Wigner spectrum for degenerate parametric amplifier}

\begin{figure}
\includegraphics[width=0.4\textwidth]{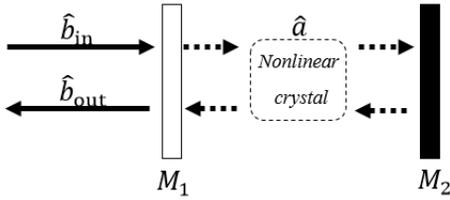}
\caption{\label{dpa} A DPA consists of a classically pumped nonlinear crystal in the Fabry-Perot cavity. Black rectangle ($M_2$) denotes a fully reflecting mirror, while white rectangle ($M_1$) denotes a partially transmitting mirror.}
\end{figure}

\begin{figure}
\includegraphics[width=0.8\textwidth]{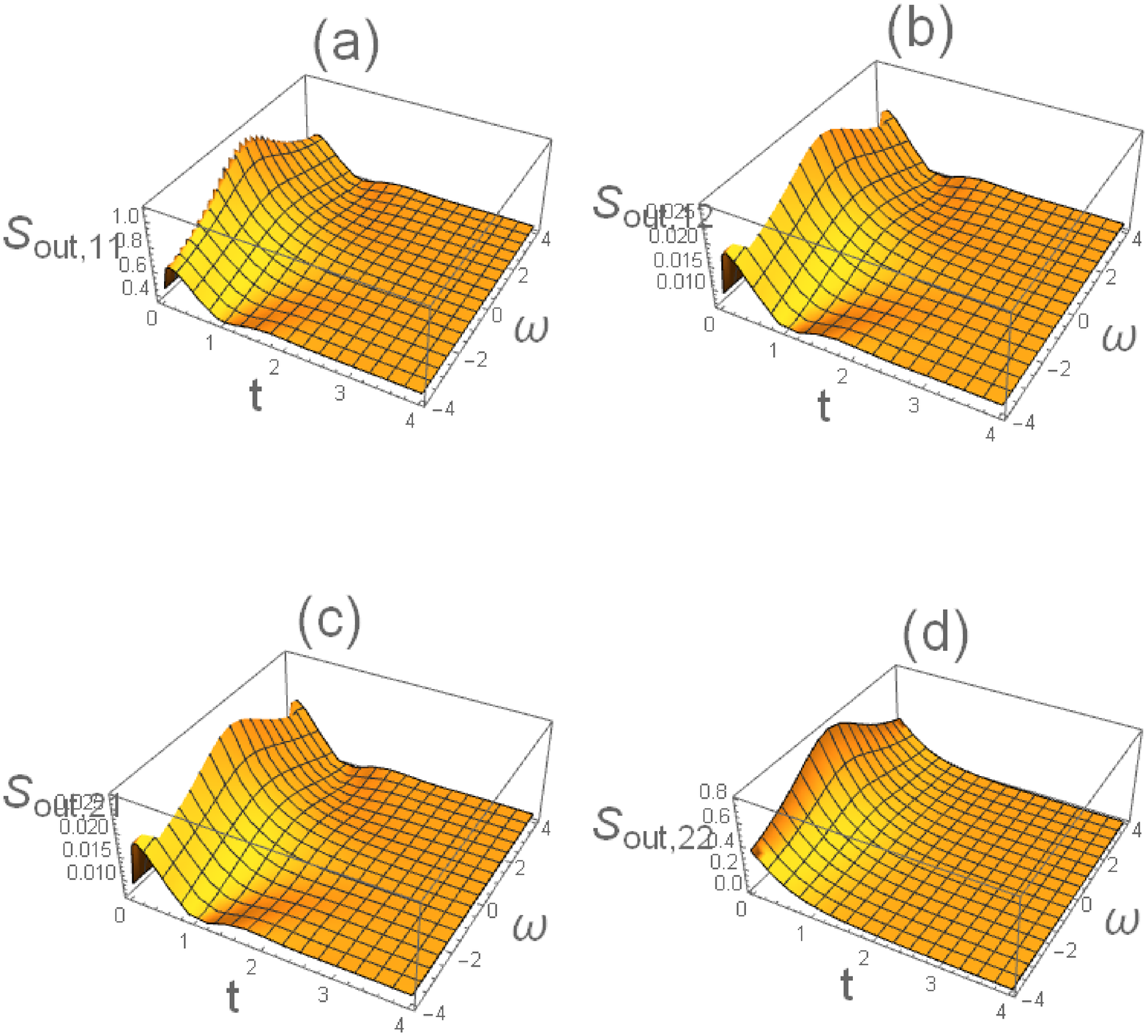}
\caption{\label{fig_30}(Color online) The output Wigner spectrum with $\epsilon=1$, $\gamma=2$ and decay rate $\kappa=100$. If we compare the four parts in one figure, it can be seen that the amplitudes in 1-by-2 and 2-by-1 entries are almost 0 (the corresponding amplitudes are less than 0.025). Thus, the output Wigner spectrum would be similar with the input when decay rate $\kappa$ is large enough although DPA is non-passive.}
\end{figure}

A degenerate parametric amplifier (DPA) is an open oscillator that is able to amplify a quadrature of the cavity mode and produce squeezed output fields, see Fig \ref{dpa} \cite[Chapter 6.3]{gardiner2004quantum}, \cite[Chapter 7.6]{walls2007quantum}, \cite[Chapter 6.3]{bachor2004guide}, \cite{nurdin2009network}. A model for a DPA is \cite{gardiner2004quantum,zhang2013response}
\numparts\begin{eqnarray}\label{42}
\left[
  \begin{array}{c}
    \dot{\hat{a}}(t) \\
    \dot{\hat{a}}^\dag(t) \\
  \end{array}
\right]
&=&-\frac{1}{2}\left[
                                 \begin{array}{cc}
                                   \kappa & -\epsilon \\
                                   -\epsilon & \kappa \\
                                 \end{array}
                               \right]\left[
                                        \begin{array}{c}
                                          \hat{a}(t) \\
                                          \hat{a}^\dag(t) \\
                                        \end{array}
                                      \right]
                               -\sqrt{\kappa}\left[
                                                                  \begin{array}{c}
                                                                    \hat{b}_{\mathrm{in}}(t) \\
                                                                    \hat{b}_{\mathrm{in}}^\dag(t) \\
                                                                  \end{array}
                                                                \right]
                               ,\\
\hat{b}_{\rm{out}}(t)&=&\sqrt{\kappa}\hat{a}(t)+\hat{b}_{\mathrm{in}}(t),~~(0<\epsilon<\kappa).
\end{eqnarray}\endnumparts
Driven by a single-photon Fock state, the steady output state is no longer a single-photon state because the DPA has pump and the system is not passive any more. The steady output state belongs to the class of photon-Gaussian states which is defined in \cite{zhang2013response}. Let the single-photon input Fock state $|1_\nu\rangle$ be that defined in \eref{31}. The output covariance function is
\begin{equation}\label{43}
R_{\rm{out}}(t,r)=\left[
               \begin{array}{cc}
                 \chi_{11}(t,r) & \chi_{12}(t,r) \\
                 \chi_{21}(t,r) & \chi_{22}(t,r) \\
               \end{array}
             \right]+
\Delta(\xi_{\rm{out}}^-(t),\xi_{\rm{out}}^+(t))\Delta(\xi_{\rm{out}}^-(r),\xi_{\rm{out}}^+(r))^\dag,
\end{equation}
whose Wigner spectrum is
\begin{equation}\label{44}
S_{\rm{out}}(t,\omega)=\left[
               \begin{array}{cc}
                 S_{\rm{out,11}}(t,\omega) & S_{\rm{out,12}}(t,\omega) \\
                 S_{\rm{out,21}}(t,\omega) & S_{\rm{out,22}}(t,\omega) \\
               \end{array}
             \right].
\end{equation}
Here, the explicit forms of output covariance function $R_{\rm{out}}(t,r)$ and Wigner spectrum $S_{\rm{out}}(t,\omega)$ are given in \ref{WignerDPA}. Similar with the cavity case, if we let decay rate $\kappa\rightarrow\infty$, \eref{41} also holds for the DPA case, which is consistent with the simulation result in Fig.~\ref{fig_30}.

\begin{figure}
\includegraphics[width=0.8\textwidth]{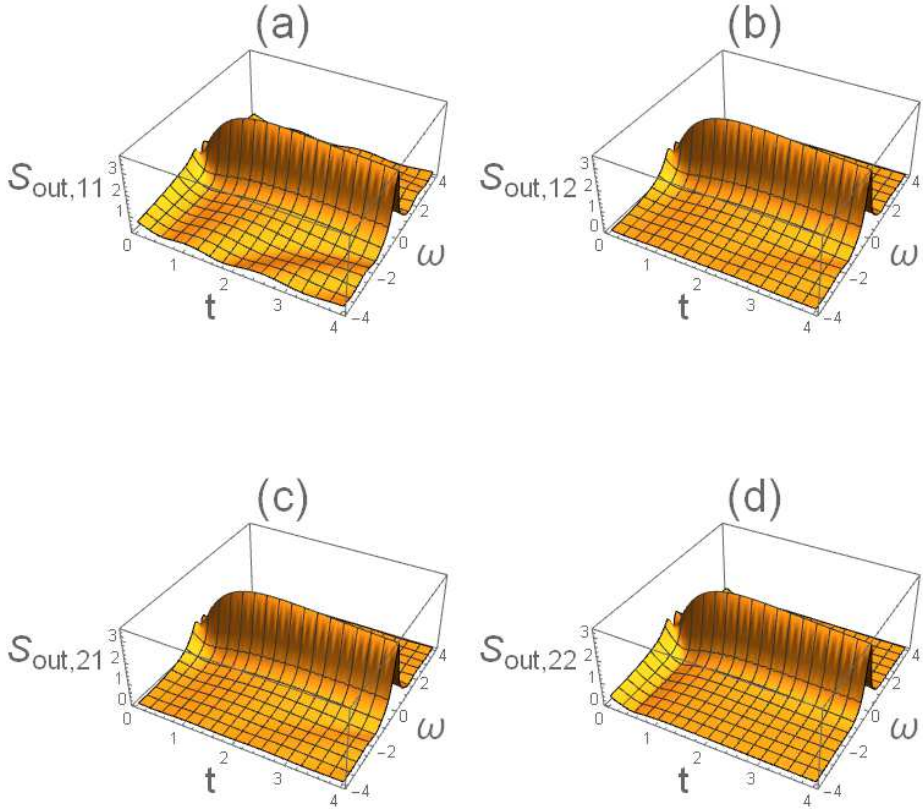}
\caption{\label{fig_27}(Color online) The output Wigner spectrum with $\epsilon=1$, $\gamma=2$ and decay rate $\kappa=1.5$. Compared with the passive system (optical cavity), the off-diagonal entries are non-zero and this output Wigner spectrum is much different since DPA is an active system. And the decay rate $\kappa$ must be greater than $\epsilon$ to make the system stable.}
\end{figure}

\begin{figure}
\includegraphics[width=0.8\textwidth]{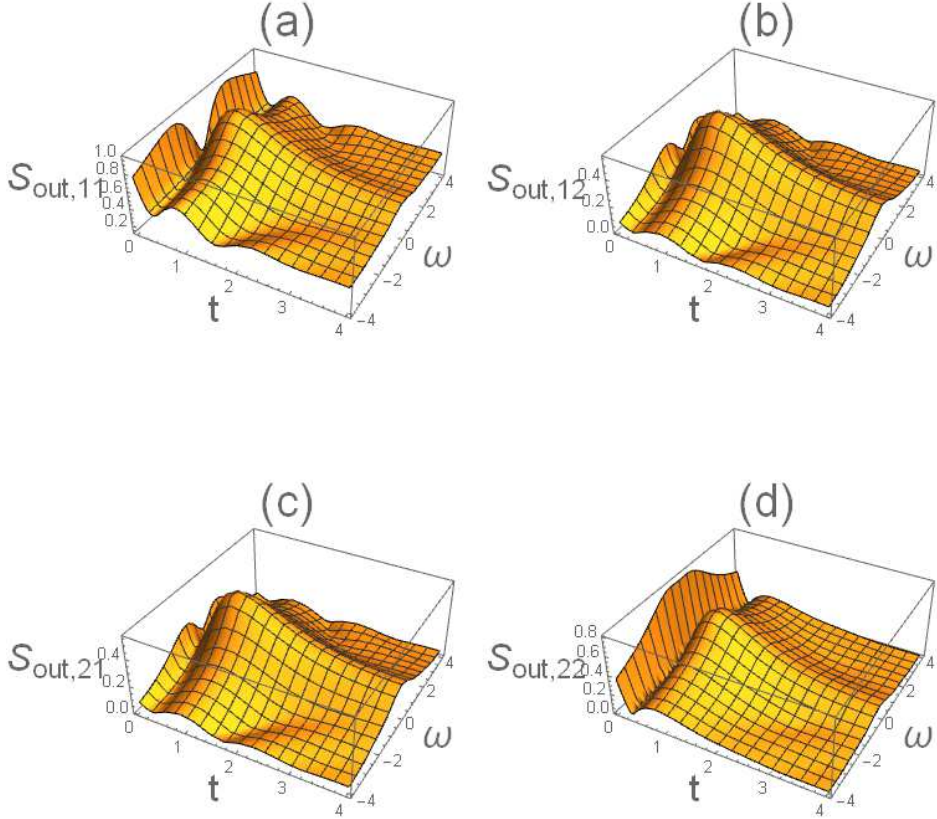}
\caption{\label{fig_29}(Color online) The output Wigner spectrum with $\epsilon=1$, $\gamma=2$ and decay rate $\kappa=4$. The output Wigner spectrum becomes non-monotonic with a larger decay rate $\kappa$. Compared with Fig. 4, it can be seen that the 1-by-1 and 2-by-2 entries converge to 0 more slowly with same decay rate $\kappa=4$. What's more, the off-diagonal entries cannot be ignored since the corresponding amplitudes are close to 0.4.}
\end{figure}

In the following we fix $\epsilon=1$, $\gamma=2$. The input Wigner spectrum is as same as the optical cavity case in Fig.~\ref{fig_15}. Figs.~\ref{fig_27}-\ref{fig_30} are simulation results for different decay rates $\kappa$, where $S_{\rm{out,11}}(t,\omega)$, $S_{\rm{out,12}}(t,\omega)$, $S_{\rm{out,21}}(t,\omega)$, $S_{\rm{out,22}}(t,\omega)$ are the entries for the output Wigner spectrum in~\eref{44} respectively. Compared with the cavity case, there exists non-zero off-diagonal parts since DPA is a non-passive system. Moreover, it can be seen clearly from Figs.~\ref{fig_27} and \ref{fig_29} that the photon-Gaussian state is significantly different from the single-photon state. A photon-Gaussian state is obtained by driving a DPA with a single-photon state, \cite{zhang2013response}.  Intuitively, a photon-Gaussian state is of the form $\mathbf{\hat{B}}^\dag(\eta)|\alpha\rangle$ in which $\eta$ is a pulse shape and $|\alpha\rangle$ is a coherent state. Clearly, when $|\alpha\rangle=|0\rangle$, we get a single-photon Fock state.

An optical cavity is a passive system while a DPA is not. By comparing figures for the cavity case and the DPA case, it can be seen that the Wigner spectrum is able to demonstrate such fundamental difference very clearly in terms of the statistical characterization of the input-output relation.

\section{Photon pulse shape engineering}\label{Pe}

\begin{figure}
\includegraphics[width=0.4\textwidth]{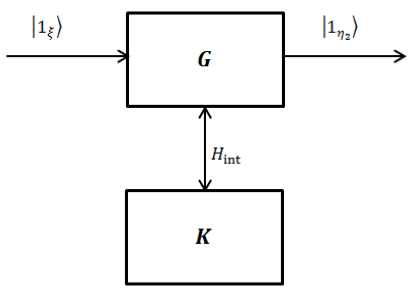}
\caption{\label{fig_33} Directly coupled system $G\bowtie K$.}
\end{figure}

\begin{figure}
\includegraphics[width=0.4\textwidth]{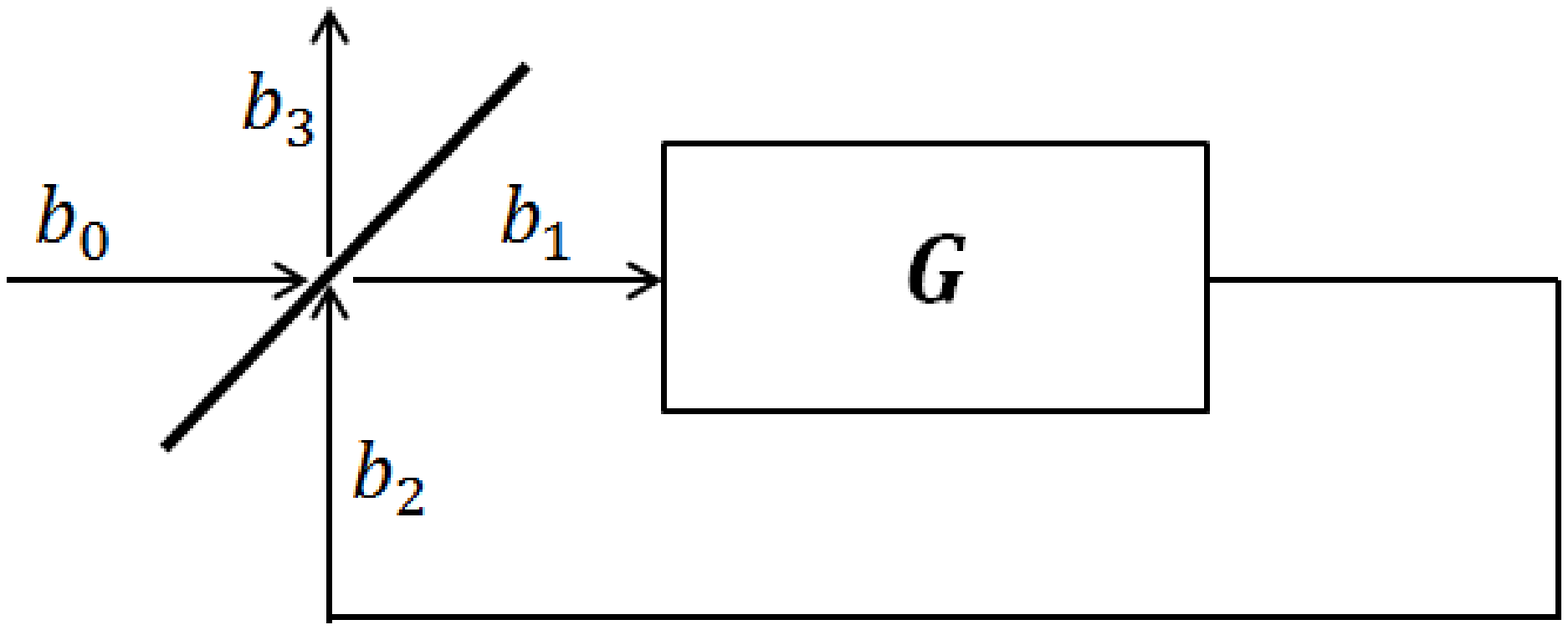}
\caption{\label{fig_34} Linear quantum feedback network consisting of a beamsplitter.}
\end{figure}

In this section, we will discuss how to engineer photon pulse shapes by means of coherent control methods, namely direct coupling (Fig.~\ref{fig_33})  and coherent feedback (Fig.~\ref{fig_34}).

\begin{figure}
\includegraphics[width=0.4\textwidth]{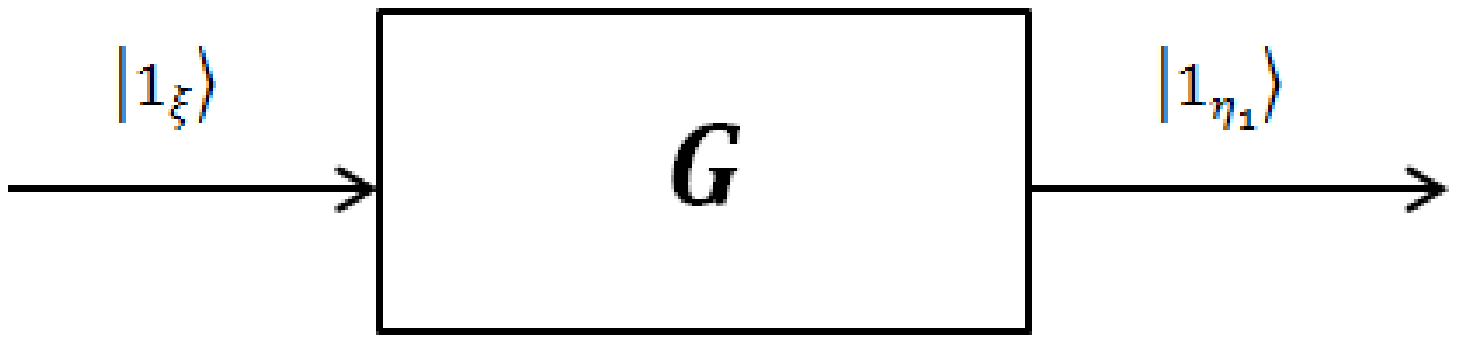}
\caption{\label{fig_32} The original system $G$.}
\end{figure}

\subsection{Direct couplings}

In Fig.~\ref{fig_33}, two independent systems $G$ and $K$ may interact by exchanging energy. This energy exchange can be described by an interaction Hamiltonian $\hat{H}_{\rm{int}}$ with the form
\begin{equation}
\hat{H}_{\rm{int}}=\hat{X}^\dag_1\hat{X}_2+\hat{X}_1\hat{X}^\dag_2,
\end{equation}
where $\hat{X}_1$ and $\hat{X}_2$ are operators on system $G$ and $K$ respectively. We can denote the directly coupled system by $G\bowtie K$, see \cite{wiseman1994all,zhang2011direct}.

Quantum markovian systems can be conveniently described by the triple $(\hat{S},\hat{L},\hat{H})$ formalism, in which $\hat{S}$ is the scattering operator matrix, $\hat{L}$ is the coupling between the system and its environment, and $\hat{H}$ denotes the initial system Hamiltonian, see \cite{gough2009series,gough2010squeezing,zhang2012quantum}.

Fig.~\ref{fig_32} is an optical cavity with the following parameters,
\begin{equation}
G=(1, \sqrt{\kappa}\hat{a}_1, \omega_1\hat{a}^\dag_1\hat{a}_1),
\end{equation}
where $\kappa$ is the system decay rate and $\omega_1$ denotes the de-tuning for system $G$.
$|1_{\xi}\rangle$ is the single-photon input Fock state and $|1_{\eta_1}\rangle$ is the output state. In Fig.~\ref{fig_33}, the system $G$ is directly coupled with another quantum system $K$ with parameters
\begin{equation}
K=(-, -, \omega_2\hat{a}^\dag_2\hat{a}_2),
\end{equation}
where $\omega_2$ denotes the de-tuning for system $K$. In this case, the output state is described by $|1_{\eta_2}\rangle$. Alternatively, we may use a beamsplitter to form a coherent feedback system, see Fig.~\ref{fig_34}. In the following, we will derive the explicit forms of output pulse shapes in the frequency domain.

\subsection{Photon shape synthesis}

Let the pulse shape of a single-photon input Fock state $|1_{\xi}\rangle$ be
\begin{eqnarray}\label{45}
\xi(t)=\left\{\begin{array}{cc}
                 \sqrt{2\beta}e^{-\beta t}, & t\geq0, \\
                 0, & t<0,
               \end{array}
\right.
\end{eqnarray}
where $\beta$ is the damping rate. By the Fourier transform, we can get the input pulse shape in the frequency domain
\begin{equation}\label{46}
\xi[\omega]=\frac{\sqrt{2\beta}}{i\omega+\beta}.
\end{equation}
The transfer function for the original system $G$ is given by
\begin{equation}\label{47}
G_1[\omega]=1-\displaystyle\frac{\kappa}{i\omega+i\omega_1+\frac{\kappa}{2}},
\end{equation}
and the output pulse shape in the frequency domain is
\begin{equation}\label{48}
\eta_1[\omega]=\left(1-\frac{\kappa}{i\omega+i\omega_1+\frac{\kappa}{2}}\right)\xi[\omega].
\end{equation}

Secondly, for the directly coupled system in Fig.~\ref{fig_33}, we assume that $\hat{X}_1=\alpha \hat{a}_1$, $\alpha\in\mathbb{C}$ and $\hat{X}_2=\hat{a}_2$. The interaction Hamiltonian is given by
\begin{equation}\label{49}
\hat{H}_{\rm{int}}=\bar{\alpha}\hat{a}^\dag_1\hat{a}_2+\alpha \hat{a}_1\hat{a}^\dag_2.
\end{equation}
Then the Hamiltonian for the whole system $G\bowtie K$ is
\begin{equation}\label{50}
\hat{H}=\hat{H}_1+\hat{H}_{\rm{int}}+\hat{H}_2,
\end{equation}
where $\hat{H}_1=\omega_1\hat{a}^\dag_1\hat{a}_1$, $\hat{H}_2=\omega_2\hat{a}^\dag_2\hat{a}_2$.

We get the  pulse shape of the output single-photon Fock state for the system $G\bowtie K$, which is
\begin{equation}\label{51}
\eta_2[\omega]=\frac{-\displaystyle\frac{\kappa}{2}(\omega+\omega_2)i-(\omega+\omega_1)(\omega+\omega_2)+|\alpha|^2}
{\displaystyle\frac{\kappa}{2}(\omega+\omega_2)i-(\omega+\omega_1)(\omega+\omega_2)+|\alpha|^2}\xi[\omega].
\end{equation}

Finally, in Fig.~\ref{fig_34}, let the beamsplitter be
\begin{equation}
S=\left[
      \begin{array}{cc}
        \sqrt{\gamma} & e^{-i\phi}\sqrt{1-\gamma} \\
        -e^{i\phi}\sqrt{1-\gamma} & \sqrt{\gamma} \\
      \end{array}
    \right],  \ 0<\gamma<1,
\end{equation}
and the input field $b_0$ be in the single-photon Fock state  $|1_{\xi}\rangle$. We can get the pulse shape for the output field $b_3$ in Fig.~\ref{fig_34}
\begin{equation}\label{52}
\eta_3[\omega]=\frac{-\displaystyle\frac{1-\sqrt{\gamma}}{1+\sqrt{\gamma}}(\omega+\omega_1)i+\frac{\kappa}{2}}
{\displaystyle\frac{1-\sqrt{\gamma}}{1+\sqrt{\gamma}}(\omega+\omega_1)i+\frac{\kappa}{2}}\xi[\omega].
\end{equation}

\subsection{Photon distribution}

For the single-photon Fock state we defined before
\begin{equation}
|1_{\xi}\rangle=\int_{-\infty}^{\infty}\hat{b}_{\mathrm{in}}^\dag(t)\xi(t)dt|0\rangle,
\end{equation}
$\hat{b}_{\mathrm{in}}^\dag(t)$ is the creation operator and $\xi(t)$ is the pulse shape which is also known as temporal wave packet. $|\xi(t)|^2$ denotes the probability of finding the photon (detection probability) in the interval $[t,t+dt)$. In this subsection, we will focus on how the system parameters change the detection probabilities in the control schemes discussed above.

By the inverse Fourier transform, we can get the output temporal wave packets
\begin{equation}\label{53}
\eta_j(t)=\frac{1}{\sqrt{2\pi}}\int_{-\infty}^{\infty}e^{i\omega t}\eta_j[\omega]d\omega,~~(j=1,2,3)
\end{equation}
where $j$ denotes the $j$-th case we discussed before.

\begin{figure}
\includegraphics[width=0.6\textwidth]{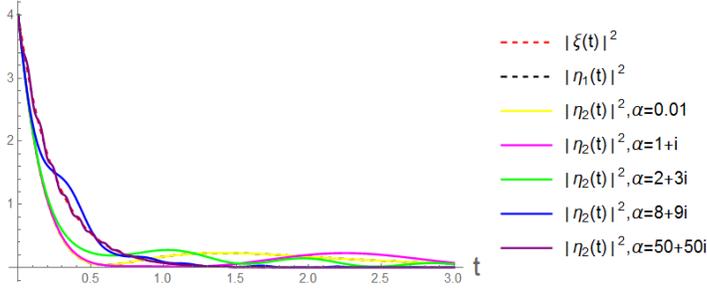}
\caption{\label{fig_41}(Color online) $|\xi(t)|^2$ denotes the detection probability of input pulse shape, $|\eta_1(t)|^2$ denotes the detection probability of output pulse shape in the case of original system (Fig.~\ref{fig_32}), $|\eta_2(t)|^2$ are the detection probabilities of output pulse shape in the directly coupled system (Fig.~\ref{fig_33}) with different parameters $\alpha$.}
\end{figure}

\begin{figure}
\includegraphics[width=0.6\textwidth]{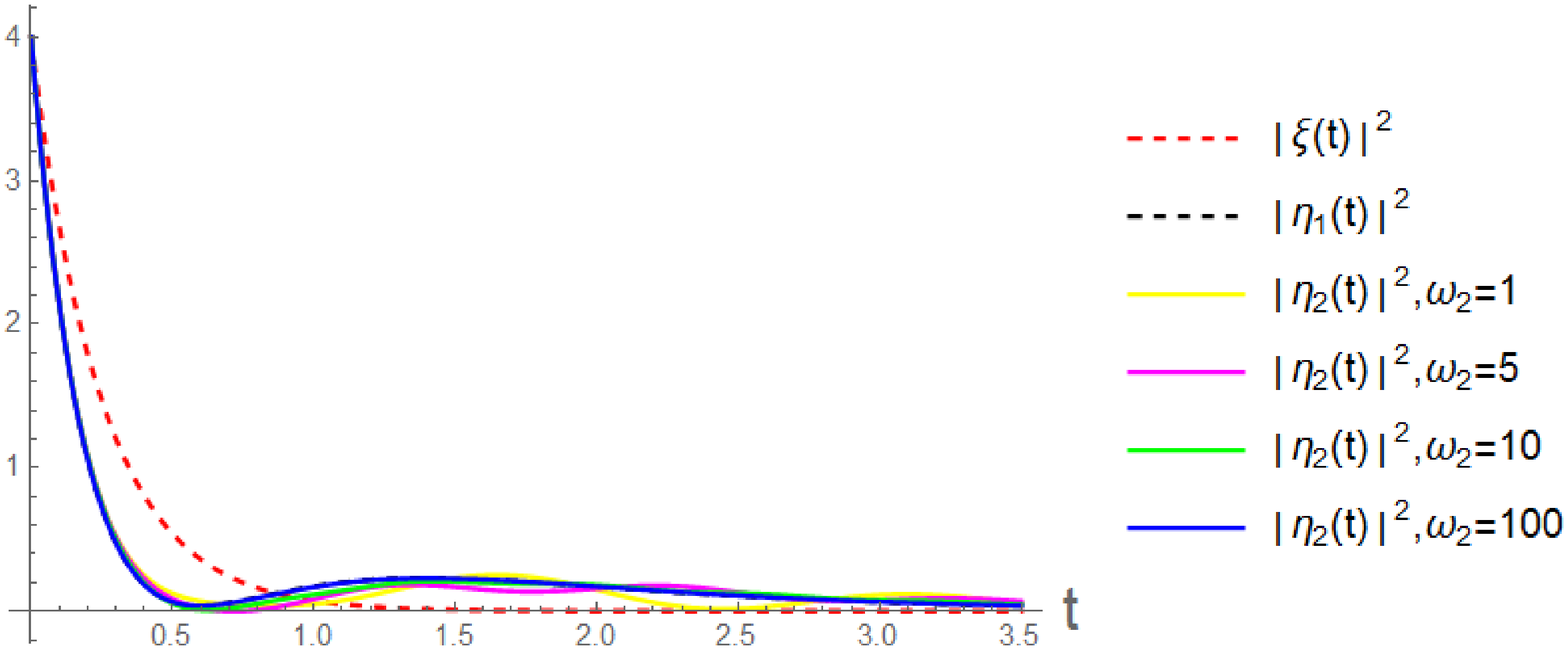}
\caption{\label{fig_42}(Color online) $|\xi(t)|^2$ denotes the detection probability of input pulse shape, $|\eta_1(t)|^2$ denotes the detection probability of output pulse shape in the case of original system (Fig.~\ref{fig_32}), $|\eta_2(t)|^2$ are the detection probabilities of output pulse shape in the directly coupled system (Fig.~\ref{fig_33}) with different parameters $\omega_2$.}
\end{figure}

\begin{figure}
\includegraphics[width=0.6\textwidth]{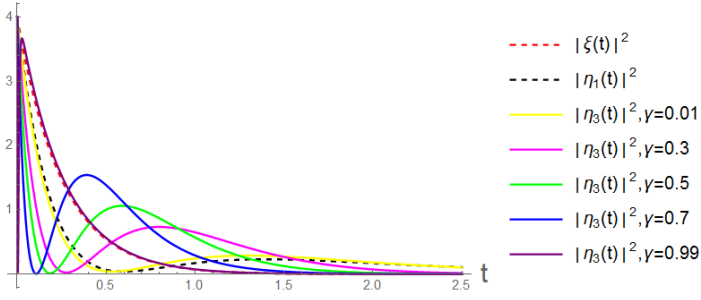}
\caption{\label{fig_43}(Color online) $|\xi(t)|^2$ denotes the detection probability of input pulse shape, $|\eta_1(t)|^2$ denotes the detection probability of output pulse shape in the case of original system (Fig.~\ref{fig_32}), $|\eta_3(t)|^2$ are the detection probabilities of output pulse shape in the linear quantum feedback network (Fig.~\ref{fig_34}) with different beamsplitter parameters $\gamma$.}
\end{figure}

For the direct coupling scheme, Fig.~\ref{fig_33}, we fix $\beta=2$, $\kappa=1$, $\omega_1=1$. Fig.~\ref{fig_41} and  Fig.~\ref{fig_42} are the detection probabilities for different $\alpha$ and $\omega_2$ respectively. For the coherent feedback control case Fig.~\ref{fig_34}, detection probabilities for different beamsplitter parameter $\gamma$ are given in Fig.~\ref{fig_43}.

\begin{figure}
\includegraphics[width=0.8\textwidth]{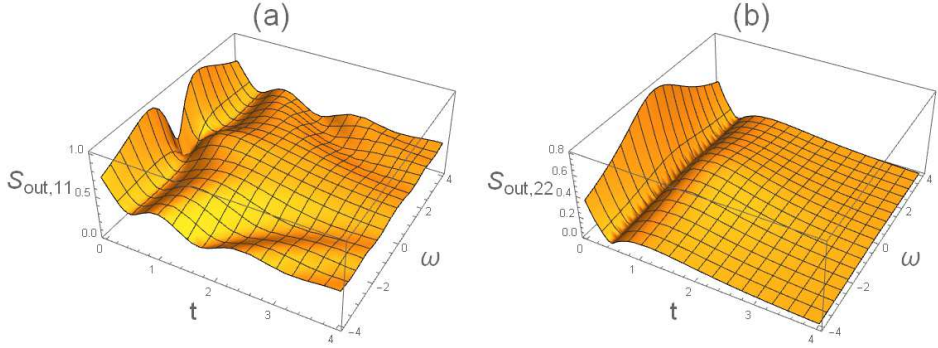}
\caption{\label{fig_b1}(Color online) The output Wigner spectrum for quantum feedback network with beamsplitter parameter $\gamma=0.01$. Since $b_3\rightarrow b_2$, $b_1\rightarrow b_0$ when $\gamma\rightarrow 0$, the feedback network should reduce to the original system without beamsplitter. This can be verified by comparison with Fig.~\ref{fig_19}.}
\end{figure}

\begin{figure}
\includegraphics[width=0.8\textwidth]{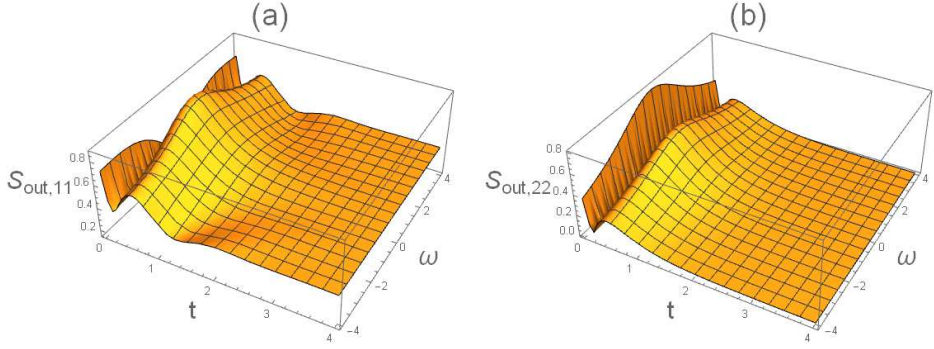}
\caption{\label{fig_b5}(Color online) The output Wigner spectrum for quantum feedback network with beamsplitter parameter $\gamma=0.5$.}
\end{figure}

\begin{figure}
\includegraphics[width=0.8\textwidth]{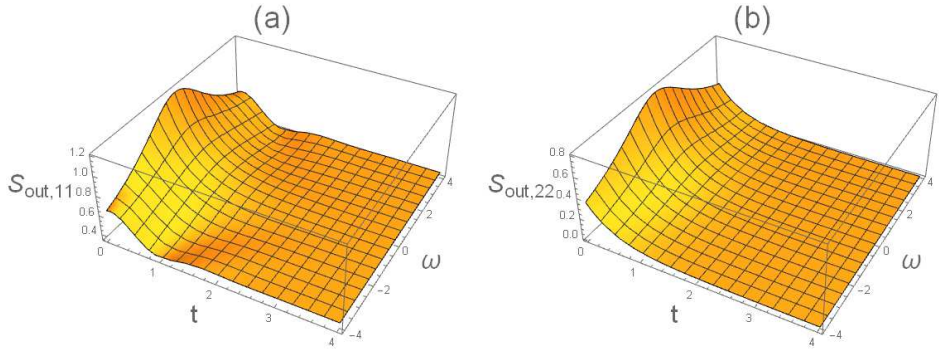}
\caption{\label{fig_b9}(Color online) The output Wigner spectrum for quantum feedback network with beamsplitter parameter $\gamma=0.99$. If $\gamma\rightarrow1$, then $b_3\rightarrow b_0$. It means that the output Wigner spectrum will be close to the input. Thus, the simulation result should be much similar with the input Wigner spectrum in Fig.~\ref{fig_15}.}
\end{figure}

By comparing these three cases, it can be easily seen that the linear quantum feedback network in Fig.~\ref{fig_34}  has much more influence on the detection probability than the directly coupled system. In addition, the changes of output Wigner spectrum with beamsplitter parameter $\gamma$ for quantum feedback network also have been analyzed. In Fig.~\ref{fig_b1} - Fig.~\ref{fig_b9}, let the decay rate of the optical cavity be $\kappa=4$ and damping rate be $\beta=2$, it can be verified that those changes are consistent with the photon distributions in Fig.~\ref{fig_43}.

On the other hand, we assume the system $G$ for the feedback network in Fig. \ref{fig_34} is a DPA with the triple $(\hat{S},\hat{L},\hat{H})$ parameters
\begin{equation}
\hat{S}_0=I,~~\hat{L}_0=\sqrt{\kappa}\hat{a},~~\hat{H}_0=\frac{i\epsilon}{4}((\hat{a}^\dag)^2-\hat{a}^2).
\end{equation}
Then the whole feedback network system parameters with beamsplitter $S$ are given by
\begin{equation}
\hat{S}_1=-I,~~\hat{L}_1=\sqrt{\frac{1+\sqrt{\gamma}}{1-\sqrt{\gamma}}\kappa}\hat{a},~~\hat{H}_1=\hat{H}_0.
\end{equation}
So the only change between the feedback network and the original system is $\kappa\rightarrow\frac{1+\sqrt{\gamma}}{1-\sqrt{\gamma}}\kappa$. There exist three cases as follows:\\
1) $\gamma=0$, the feedback network reduces to the open-loop system $G$.\\
2) $\gamma=1$, then $\hat{S}=I$, $b_3=b_0$, there is no interaction between field and system.\\
3) $0<\gamma<1$, $\frac{1+\sqrt{\gamma}}{1-\sqrt{\gamma}}\kappa>\kappa$, the decay rate is always enhanced. However, it is clear that
\begin{equation}
\displaystyle{\lim_{\gamma\rightarrow0}}\frac{1+\sqrt{\gamma}}{1-\sqrt{\gamma}}\kappa=\kappa.
\end{equation}
Therefore, by tuning the beamsplitter we can get various output single-photon states.  It is worth noting that the same feedback scheme Fig.~\ref{fig_34} has been used for optical squeezing, see theoretical \cite{GW09} and experimental \cite{IYY+12}.

\section{Conclusion}\label{Con}

In this article, the Wigner spectrum has been used to analyze the statistical properties of continuous-mode single-photon Fock states.  The Wigner spectrum is able to show the significant difference between the statistical nature of the output fields of an optical cavity and a degenerate parametric amplifier (DPA), driven by a continuous-mode  single-photon Fock state.  Several control schemes are compared for photon pulse-shaping. It has been demonstrated that the coherent feedback control scheme is very effective in photon pulse-shaping.

\bibliographystyle{iopart-num}
\bibliography{dzy}

\providecommand{\newblock}{}
\begin{thebibliography}{10}
\expandafter\ifx\csname url\endcsname\relax
  \def\url#1{{\tt #1}}\fi
\expandafter\ifx\csname urlprefix\endcsname\relax\def\urlprefix{URL }\fi
\providecommand{\eprint}[2][]{\url{#2}}

\bibitem{beveratos2002single}
Beveratos A, Brouri R, Gacoin T, Villing A, Poizat J~P and Grangier P 2002 {\em
  Physical Review Letters\/} {\bf 89} 187901

\bibitem{gisin2007quantum}
Gisin N and Thew R 2007 {\em Nature photonics\/} {\bf 1} 165--171

\bibitem{loudon2000quantum}
Loudon R 2000 {\em The quantum theory of light\/} (Oxford university press)

\bibitem{knill2001scheme}
Knill E, Laflamme R and Milburn G~J 2001 {\em nature\/} {\bf 409} 46--52

\bibitem{giovannetti2004quantum}
Giovannetti V, Lloyd S and Maccone L 2004 {\em Science\/} {\bf 306} 1330--1336

\bibitem{giovannetti2011advances}
Giovannetti V, Lloyd S and Maccone L 2011 {\em Nature Photonics\/} {\bf 5} 222

\bibitem{leroux2012unitary}
Leroux I~D, Schleier-Smith M~H, Zhang H and Vuleti{\'c} V 2012 {\em Physical
  Review A\/} {\bf 85} 013803

\bibitem{moehring2007entanglement}
Moehring D, Maunz P, Olmschenk S, Younge K, Matsukevich D, Duan L~M and Monroe
  C 2007 {\em Nature\/} {\bf 449} 68

\bibitem{kimble2008quantum}
Kimble H~J 2008 {\em Nature\/} {\bf 453} 1023

\bibitem{aghamalyan2011quantum}
Aghamalyan D and Malakyan Y 2011 {\em Physical Review A\/} {\bf 84} 042305

\bibitem{blow1990continuum}
Blow K, Loudon R, Phoenix S~J and Shepherd T 1990 {\em Physical Review A\/}
  {\bf 42} 4102

\bibitem{braunstein2005quantum}
Braunstein S~L and Van~Loock P 2005 {\em Reviews of Modern Physics\/} {\bf 77}
  513

\bibitem{furusawa2011quantum}
Furusawa A and Van~Loock P 2011 {\em Quantum teleportation and entanglement: a
  hybrid approach to optical quantum information processing\/} (John Wiley \&
  Sons)

\bibitem{brecht2015photon}
Brecht B, Reddy D~V, Silberhorn C and Raymer M~G 2015 {\em Physical Review X\/}
  {\bf 5} 041017

\bibitem{garrison2008quantum}
Garrison J and Chiao R 2008 {\em Quantum optics\/} (Oxford University Press)

\bibitem{milburn2008coherent}
Milburn G 2008 {\em The European Physical Journal Special Topics\/} {\bf 159}
  113

\bibitem{ogawa2015real}
Ogawa H, Ohdan H, Miyata K, Taguchi M, Makino K, Yonezawa H, Yoshikawa J~i and
  Furusawa A 2016 {\em Physical Review Letters\/} {\bf 116} 233602

\bibitem{WMS+11}
Wang Y, Min{\'a}{\v{r}} J, Sheridan L and Scarani V 2011 {\em Physical Review
  A\/} {\bf 83} 063842

\bibitem{gough2011quantum}
Gough J~E, James M~R and Nurdin H~I 2011 {\em 50th IEEE Conference on Decision
  and Control and European Control Conference\/}  5570--5576

\bibitem{gough2012quantum}
Gough J~E, James M~R, Nurdin H~I and Combes J 2012 {\em Physical Review A\/}
  {\bf 86} 043819

\bibitem{gough2013quantum}
Gough J~E, James M~R and Nurdin H~I 2013 {\em Quantum information processing\/}
  {\bf 12} 1469

\bibitem{baragiola2012n}
Baragiola B~Q, Cook R~L, Bra{\'n}czyk A~M and Combes J 2012 {\em Physical
  Review A\/} {\bf 86} 013811

\bibitem{SZX16}
Song H, Zhang G and Xi Z 2016 {\em SIAM Journal on Control and Optimization\/}
  {\bf 54} 1602--1632

\bibitem{carvalho2012cavity}
Carvalho A, Hush M and James M 2012 {\em Physical Review A\/} {\bf 86} 023806

\bibitem{gheri1998photon}
Gheri K~M, Ellinger K, Pellizari T and Zoller P 1998 {\em Fortschritte der
  Physik\/} {\bf 46} 401--416

\bibitem{zhang2013response}
Zhang G and James M~R 2013 {\em Automatic Control, IEEE Transactions on\/} {\bf
  58} 1221

\bibitem{yamamoto2014zero}
Yamamoto N and James M~R 2014 {\em New Journal of Physics\/} {\bf 16} 073032

\bibitem{qin2014complete}
Qin Z, Prasad A~S, Brannan T, MacRae A, Lezama A and Lvovsky A 2015 {\em Light:
  Science \& Applications\/} {\bf 4} e298

\bibitem{munro2010intracavity}
Munro W, Nemoto K and Milburn G 2010 {\em Optics Communications\/} {\bf 283}
  741--746

\bibitem{pan2014input}
Pan Y, Zhang G and James M~R 2016 {\em Automatica\/} {\bf 69} 18--23

\bibitem{gardiner2004quantum}
Gardiner C and Zoller P 2004 {\em Quantum noise: a handbook of Markovian and
  non-Markovian quantum stochastic methods with applications to quantum
  optics\/} vol~56 (Springer Science \& Business Media)

\bibitem{gough2005quantum}
Gough J~E 2005 {\em Russian Journal of Mathematical Physics\/} {\bf 10}
  142--148

\bibitem{wigner1932quantum}
Wigner E 1932 {\em Physical Review\/} {\bf 40} 749

\bibitem{ville1948theorie}
Ville J~d {\em et~al.\/} 1948 {\em Cables et transmission\/} {\bf 2} 61--74

\bibitem{sandsten2013time}
Sandsten M 2013 {\em Time-Frequency Analysis of Non-Stationary Processes\/}
  (Lund University, Centre for Mathematical Sciences)

\bibitem{yanagisawa2003transfer}
Yanagisawa M and Kimura H 2003 {\em IEEE Transactions on Automatic control\/}
  {\bf 48} 2107--2120

\bibitem{bachor2004guide}
Bachor H~A and Ralph T~C 2004 {\em A guide to experiments in quantum optics\/}
  (Wiley)

\bibitem{walls2007quantum}
Walls D~F and Milburn G~J 2007 {\em Quantum optics\/} (Springer Science \&
  Business Media)

\bibitem{nurdin2009network}
Nurdin H~I, James M~R and Doherty A~C 2009 {\em SIAM Journal on Control and
  Optimization\/} {\bf 48} 2686--2718

\bibitem{wiseman1994all}
Wiseman H~M and Milburn G~J 1994 {\em Physical review A\/} {\bf 49} 4110

\bibitem{zhang2011direct}
Zhang G and James M~R 2011 {\em Automatic Control, IEEE Transactions on\/} {\bf
  56} 1535

\bibitem{gough2009series}
Gough J and James M~R 2009 {\em Automatic Control, IEEE Transactions on\/} {\bf
  54} 2530

\bibitem{gough2010squeezing}
Gough J~E, James M and Nurdin H 2010 {\em Physical Review A\/} {\bf 81} 023804

\bibitem{zhang2012quantum}
Zhang G and James M~R 2012 {\em Chinese Science Bulletin\/} {\bf 57} 2200

\bibitem{GW09}
Gough J~E and Wildfeuer S 2009 {\em Physical Review A\/} {\bf 80} 042107

\bibitem{IYY+12}
Iida S, Yukawa M, Yonezawa H, Yamamoto N and Furusawa A 2012 {\em Automatic
  Control, IEEE Transactions on\/} {\bf 57} 2045

\end{thebibliography}

\appendix

\section{\label{WignerDPA}The explicit form of output Wigner spectrum for DPA}

If the single-photon input $|1_\nu\rangle$ has the pulse shape defined in \eref{31}, we can get the pulse shape of output state
\begin{eqnarray*}
\xi_{\rm{out}}^-(t)=\frac{(\epsilon^2+\kappa^2-4\gamma^2)\sqrt{2\gamma}}{(\kappa+\epsilon-2\gamma)(\epsilon-\kappa+2\gamma)}e^{-\gamma t}+\frac{\kappa\sqrt{2\gamma}}{\kappa+\epsilon-2\gamma}e^{(-\frac{\epsilon}{2}-\frac{\kappa}{2})t}-\frac{\kappa\sqrt{2\gamma}}{\epsilon-\kappa+2\gamma}e^{(\frac{\epsilon}{2}-\frac{\kappa}{2})t},\\
\xi_{\rm{out}}^+(t)=\frac{2\kappa\epsilon\sqrt{2\gamma}}{(\kappa+\epsilon-2\gamma)(\epsilon-\kappa+2\gamma)}e^{-\gamma t}-\frac{\kappa\sqrt{2\gamma}}{\kappa+\epsilon-2\gamma}e^{(-\frac{\epsilon}{2}-\frac{\kappa}{2})t}-\frac{\kappa\sqrt{2\gamma}}{\epsilon-\kappa+2\gamma}e^{(\frac{\epsilon}{2}-\frac{\kappa}{2})t}.
\end{eqnarray*}

Then, the output covariance function is
\begin{eqnarray*}
R_{\rm{out}}(t,r)=\left[
               \begin{array}{cc}
                 \chi_{11}(t,r) & \chi_{12}(t,r) \\
                 \chi_{21}(t,r) & \chi_{22}(t,r) \\
               \end{array}
             \right]+\Delta(\xi_{\rm{out}}^-(t),\xi_{\rm{out}}^+(t))\Delta(\xi_{\rm{out}}^-(r),\xi_{\rm{out}}^+(r))^\dag,
\end{eqnarray*}
where
\begin{eqnarray}
\chi_{11}(t,r)=&\left\{\begin{array}{cc}
                        \frac{-\kappa\epsilon}{4(\kappa+\epsilon)}e^{(-\frac{\epsilon}{2}-\frac{\kappa}{2})(t-r)}+\frac{\kappa\epsilon}{4(\kappa-\epsilon)}e^{(\frac{\epsilon}{2}-\frac{\kappa}{2})(t-r)}, & t>r, \\
                        \delta(t-r)+\frac{3\kappa\epsilon^2-2\kappa^3}{2(\kappa^2-\epsilon^2)}, & t=r, \\
                        \frac{-\kappa\epsilon}{4(\kappa+\epsilon)}e^{(-\frac{\epsilon}{2}-\frac{\kappa}{2})(r-t)}+\frac{\kappa\epsilon}{4(\kappa-\epsilon)}e^{(\frac{\epsilon}{2}-\frac{\kappa}{2})(r-t)}, & t<r,
                      \end{array}
\right.\\
\chi_{12}(t,r)=&\left\{\begin{array}{cc}
                        \frac{\kappa\epsilon}{4(\kappa+\epsilon)}e^{(-\frac{\epsilon}{2}-\frac{\kappa}{2})(t-r)}+\frac{\kappa\epsilon}{4(\kappa-\epsilon)}e^{(\frac{\epsilon}{2}-\frac{\kappa}{2})(t-r)}, & t>r, \\
                        \frac{\kappa^2\epsilon}{2(\kappa^2-\epsilon^2)}, & t=r, \\
                        \frac{\kappa\epsilon}{4(\kappa+\epsilon)}e^{(-\frac{\epsilon}{2}-\frac{\kappa}{2})(r-t)}+\frac{\kappa\epsilon}{4(\kappa-\epsilon)}e^{(\frac{\epsilon}{2}-\frac{\kappa}{2})(r-t)}, & t<r,
                      \end{array}
\right.\\
\chi_{21}(t,r)=&\chi_{12}(t,r),\\
\chi_{22}(t,r)=&\left\{\begin{array}{cc}
                        \frac{-\kappa\epsilon}{4(\kappa+\epsilon)}e^{(-\frac{\epsilon}{2}-\frac{\kappa}{2})(t-r)}+\frac{\kappa\epsilon}{4(\kappa-\epsilon)}e^{(\frac{\epsilon}{2}-\frac{\kappa}{2})(t-r)}, & t>r, \\
                        \frac{\kappa\epsilon^2}{2(\kappa^2-\epsilon^2)}, & t=r, \\
                        \frac{-\kappa\epsilon}{4(\kappa+\epsilon)}e^{(-\frac{\epsilon}{2}-\frac{\kappa}{2})(r-t)}+\frac{\kappa\epsilon}{4(\kappa-\epsilon)}e^{(\frac{\epsilon}{2}-\frac{\kappa}{2})(r-t)}, & t<r.
                      \end{array}
\right.
\end{eqnarray}

Thus, the Wigner spectrum of output covariance function for the DPA is
\begin{eqnarray}
S_{\rm{out}}(t,\omega)=\left[
               \begin{array}{cc}
                 S_{\rm{out,11}}(t,\omega) & S_{\rm{out,12}}(t,\omega) \\
                 S_{\rm{out,21}}(t,\omega) & S_{\rm{out,22}}(t,\omega) \\
               \end{array}
             \right],
\end{eqnarray}
where
\begin{eqnarray*}
&S_{\rm{out,11}}(t,\omega)=\frac{1}{\sqrt{2\pi}}\times\Big\{\frac{-\kappa\epsilon}{4(\kappa+\epsilon)(\frac{\kappa}{2}+\frac{\epsilon}{2}-i\omega)}e^{-i\omega t}+\frac{\kappa\epsilon}{4(\kappa-\epsilon)(\frac{\kappa}{2}-\frac{\epsilon}{2}-i\omega)}e^{-i\omega t}\\
&+\frac{-\kappa\epsilon}{4(\kappa+\epsilon)(\frac{\kappa}{2}+\frac{\epsilon}{2}+i\omega)}e^{-i\omega t}+\frac{\kappa\epsilon}{4(\kappa-\epsilon)(\frac{\kappa}{2}-\frac{\epsilon}{2}+i\omega)}e^{-i\omega t}+e^{-i\omega t}\\
&+\xi_{\rm{out}}^-(t)[\frac{(\epsilon^2+\kappa^2-4\gamma^2)\sqrt{2\gamma}}{(\epsilon+\kappa-2\gamma)(\epsilon-\kappa+2\gamma)(\gamma+i\omega)}+\frac{\kappa\sqrt{2\gamma}}{(\kappa+\epsilon-2\gamma)(\frac{\kappa}{2}+\frac{\epsilon}{2}+i\omega)}\\
&+\frac{\kappa\sqrt{2\gamma}}{(\kappa-\epsilon-2\gamma)(\frac{\kappa}{2}-\frac{\epsilon}{2}+i\omega)}]
+\xi_{\rm{out}}^+(t)[\frac{2\kappa\epsilon\sqrt{2\gamma}}{(\epsilon+\kappa-2\gamma)(\epsilon-\kappa+2\gamma)(\gamma+i\omega)}\\
&-\frac{\kappa\sqrt{2\gamma}}{(\kappa+\epsilon-2\gamma)(\frac{\kappa}{2}+\frac{\epsilon}{2}+i\omega)}+\frac{\kappa\sqrt{2\gamma}}{(\kappa-\epsilon-2\gamma)(\frac{\kappa}{2}-\frac{\epsilon}{2}+i\omega)}]\Big\},\\
&S_{\rm{out,12}}(t,\omega)=\frac{1}{\sqrt{2\pi}}\times\Big\{\frac{\kappa\epsilon}{4(\kappa+\epsilon)(\frac{\epsilon}{2}+\frac{\kappa}{2}-i\omega)}e^{-i\omega t}+\frac{\kappa\epsilon}{4(\kappa-\epsilon)(\frac{\kappa}{2}-\frac{\epsilon}{2}-i\omega)}e^{-i\omega t}\\
&+\frac{\kappa\epsilon}{4(\kappa+\epsilon)(\frac{\kappa}{2}+\frac{\epsilon}{2}+i\omega)}e^{-i\omega t}
+\frac{\kappa\epsilon}{4(\kappa-\epsilon)(\frac{\kappa}{2}-\frac{\epsilon}{2}+i\omega)}e^{-i\omega t}\\
&+\xi_{\rm{out}}^-(t)[\frac{2\kappa\epsilon\sqrt{2\gamma}}{(\epsilon+\kappa-2\gamma)(\epsilon-\kappa+2\gamma)(\gamma+i\omega)}-\frac{\kappa\sqrt{2\gamma}}{(\kappa+\epsilon-2\gamma)(\frac{\kappa}{2}+\frac{\epsilon}{2}+i\omega)}\\
&+\frac{\kappa\sqrt{2\gamma}}{(\kappa-\epsilon-2\gamma)(\frac{\kappa}{2}-\frac{\epsilon}{2}+i\omega)}]
+\xi_{\rm{out}}^+(t)[\frac{(\epsilon^2+\kappa^2-4\gamma^2)\sqrt{2\gamma}}{(\epsilon+\kappa-2\gamma)(\epsilon-\kappa+2\gamma)(\gamma+i\omega)}\\
&+\frac{\kappa\sqrt{2\gamma}}{(\kappa+\epsilon-2\gamma)(\frac{\kappa}{2}+\frac{\epsilon}{2}+i\omega)}+\frac{\kappa\sqrt{2\gamma}}{(\kappa-\epsilon-2\gamma)(\frac{\kappa}{2}-\frac{\epsilon}{2}+i\omega)}]\Big\},\\
&S_{\rm{out,21}}(t,\omega)=S_{\rm{out,12}}(t,\omega),~~S_{\rm{out,22}}(t,\omega)=S_{\rm{out,11}}(t,\omega)-\frac{1}{\sqrt{2\pi}}e^{-i\omega t}.
\end{eqnarray*}

%
%
%
%
%
%
%

\end{document}